\documentclass[12pt,a4paper]{article}
\usepackage{amssymb,amsmath,euscript,latexsym}

\def\<{\langle}
\def\>{\rangle}

\newcommand{\md}[1]{~(\text{mod}~#1)}

\begin{document}
\title{
\textsc{\bf On Classical and Quantum Cryptography}
 \footnote{Lectures at the
Volterra--CIRM International School "Quantum Computer and  Quantum
Information", Trento, Italy, July 25--31, 2001. }\\
$~$\\}
\author{
\textsf{I.V.~Volovich}
\\
\emph{Steklov Mathematical Institute, Russian Academy of Science}\\
\emph{Gubkin st. 8, GSP-1, 117966 Moscow, Russia }\\
\emph{email: volovich@mi.ras.ru}
\\~\\
\textsf{and}\\~\\
\textsf{Ya.I.~Volovich}
\\
\emph{Physics Department, Moscow State University}\\
\emph{Vorobievi Gori, 119899 Moscow, Russia}\\
\emph{email: yaroslav\_v@mail.ru}
}

\date {~}
\maketitle
\thispagestyle{empty}

\begin{abstract}
Lectures on  classical and  quantum cryptography. Contents:
Private key cryptosystems. Elements of number theory.
Public key cryptography and RSA cryptosystem.
Shannon`s entropy and mutual information. Entropic uncertainty 
relations. The no cloning theorem. The BB84 quantum cryptographic
protocol. Security proofs. Bell`s theorem. 
The EPRBE quantum cryptographic protocol.
\end{abstract}
\newpage
\tableofcontents

\section{Introduction}
Cryptography is the art of code-making, code-breaking and secure
communication. It has a long history of military, diplomatic and
commercial applications dating back to ancient societies.
In these lectures an introduction to basic notions
of classical and  quantum cryptography is given.

A well known example of cryptosystem is the Caesar cipher. Julius
Caesar allegedly used a simple letter substitution method. Each
letter of Caesar`s message was replaced by the letter that
followed it alphabetically by 3 places. This method is called the
Caesar cipher. The size of the shift (3 in this example) should be
kept secret. It is called {\it the key} of the cryptosystem. It is
an example of the traditional cryptosystem.  It is also called the
{\it private key cryptography}. Anyone who knew the enciphering
key can decipher the message. Mathematical theory of classical
cryptography has been developed by C. Shannon.

There is a problem in the
private key cryptography which is called the {\it problem of  key
distribution}. To establish the key, two users must use a very
secure channel. In {\it classical world} an eavesdropper in
principle can monitor the channel without the legitimate users
being aware that an eavesdropping has taken place.

 In 1976 W. Diffie and M. Hellman \cite{DH} discovered a new type
of cryptosystem and invented {\it public key cryptography}. In
this method the problem of key distribution was solved. A public
key cryptosystem has the property that someone who knows only how
to encipher cannot use the enciphering key to find the deciphering
key without a prohibitively lengthy computation. The best-known
public key cryptosystem, RSA \cite{RSA}, is widely used in Internet
and other business. The system relies on the difficulty of
factoring large integers.

In the 1970`s S. Wiesner \cite{Wie}
and C.H. Bennett and G. Brassard \cite{BB}
(their method is the called BB84 protocol)
have proposed the idea of quantum cryptography.
They used the sending of single quantum particles. The method
of quantum cryptography also can solve the key distribution
problem. Moreover it can detect the presence of an eavesdropper.
In 1991 A. Ekert \cite{Eke} proposed to use in quantum cryptography
the phenomena of entanglement and Bell`s inequalities.

Experimental quantum key distribution was demonstrated
for the first time in 1989 and since then 
tremendous progress has been made.
Several groups have shown that quantum key distribution
is possible, even outside the laboratory. In particular it was reported 
the creation of a key over the distance of several
dozens kilometers
\cite{GRTZ}.

First we discuss Caesar`s cryptosystem and then in Sect.3 elements of
number theory needed for cryptography are discussed. 
In Section 4 the public
key distribution and the RSA cryptosystem is considered.
The BB84 quantum cryptographic protocol is discussed in Sect.8.
Some useful notions of the mutual information and Shannon`s entropy
are included and proofs of security of the protocol is discussed.
In Sect 9. the Einstein-Podolsky-Rosen-Bell-Ekert
(EPRBE) quantum cryptographic protocol is considered.
The security of the protocol is based on Bell`s theorem
describing nonlocal properties of entangled states.
The importance of consideration of entangled states
in space and time is stressed.
A modification of Bell`s equation which includes
the spacetime variables is given and
 the problem of security of the EPRBE protocol in real spacetime
 is discussed.


\section{Private Key Cryptosystems}

Cryptography is the art of sending messages in disguised form. We shall use
the following notions.

{\bf Alphabet} - a set of letters.

{\bf Plaintext} - the message we want to send.

{\bf Ciphertext} - the disguised message.

The plaintext and ciphertext are broken up into {\it message units}. A
message unit might be a single letter, a pair of letters or a block of $k$
letters.

An {\it enciphering} transformation is a function $f$ from the set
$\EuScript{P}$ of all possible plaintext message units to the set
$\EuScript{C}$ of all possible ciphertext units. We assume that $f$ is
a $1-\mbox{to}-1$ correspondence. $f: \EuScript{P}\to\EuScript{C}$. The {\it
deciphering} transformation is the map $f^{-1}$ which goes back and recovers
the plaintext from the ciphertext. Schematically one has the diagram
$$
\EuScript{P} \xrightarrow{f} 
\EuScript{C}\xrightarrow{f^{-1}} \EuScript{P}
$$

Any such set-up is called a {\it cryptosystem}.

\subsection{Julius Caesar's cryptosystem}

Let us discuss the Caesar cryptosystem in more detail.
Suppose we use the 26-letter Latin alphabet $A,B,\ldots,Z$ with numerical
equivalents $0,1,\ldots,25$. Let the letter $x\in\{0,1,\ldots,25\}$ stands
for a plaintext message unit. Define a function
$$
f: \{0,\ldots,25\}\to\{0,\ldots,25\}
$$
by the rule
$$
f(x)=
\begin{cases}
x+3, & \text{if $x < 23$}\\
x+3-26=x-23, & \text{if $x \geq 23$}
\end{cases}
$$
In other words $f(x)\equiv x+3\md{26}$.

To decipher a message one subtracts $3$ modulo $26$.

~

{\bf Exercise.} 
According to the Caesar`s cryptosystem the word "COLD"  reads
"FROG".

~

More generally consider the congruence (see Sect. 3 about
the properties of  congruences)
$$f(x)=x+b\md{N}$$
i.e.
$$
\begin{cases}
x+b, & \text{if $x < N-b$}\\
x-(N-b)=x+b-N, & \text{if $x \geq N-b$}
\end{cases}
$$
In the case of Caesar`s cryptosystem $N=26$, $b=3$.
To decipher the message one subtracts $b$ modulo $N$.

We could use a more general {\it affine map}, i.e. $f(x)=ax+b\md{N}$.
To decipher a message $y=ax+b\md{N}$ one solves for $x$ in terms of $y$
obtaining
$$
x=a'y+b'\md{N}
$$
where $a'$ is the inverse of $a$ modulo $N$ and $b'=-a^{-1}b\md {N}$.
Assume $a$ is relatively prime to $N$, then there exists $a^{-1}$
(see Sect.3).

In this example the enciphering function $f$ depends
upon the choice  of parameters $a$ and $b$. The values of
parameters are called the {\it enciphering key} $K_E=(a,b)$. In order to
compute $f^{-1}$ (decipher) we need a {\it deciphering key} $K_D$.
In our example $K_D=(a',b')$ where $a'=a^{-1}\md{N}$
and $b'=-a^{-1}b\md {N}$.

\subsection{Symmetric Cryptosystems - DES and GOST}

Suppose that the algorithm of the cryptosystem is publicly known but the
keys are kept in secret. It is a {\it private key cryptography}. Examples
of such cryptosystems are Data Encryption Standard (DES),  with 56-bit
private key (USA, 1980) and a more secure GOST-28147-89 which uses 256-bit
key (Russia, 1989). In such cryptosystems anyone who knows an enciphering
key can determine the deciphering key. Such cryptosystems are called {\it
symmetric cryptosystems}.

\section{Elements of Number Theory}\label{Numb}

In this section we  collect some relevant material from number theory
\cite{Vin}.

~

\textbf{Euclid`s Algorithm. }Given two integers $a$ and $b,$ not
both
zero, the \textit{greatest common divisor }of $a$ and $b,$ denoted \textit{%
g.c.d.}$(a,b\mathit{)}$\textit{\ }is the biggest integer\textit{\
}$d$ dividing both $a$ and $b.$  For example,
\textit{g.c.d.(}$9,12$\textit{)}$=3.$

There is the well known \textit{Euclid`s algorithm} of finding the
greatest common divisor. It proceeds as follows.

Find \textit{g.c.d.}$(a,b\mathit{)}$ where $a>b>0.$

1) Divide $b$ into $a$ and write down the quotient $q_{1}$ and the
remainder $r_{1}:$
\[
a=q_{1}b+r_{1}, \quad 0 < r_1 <b,
\]
2) Next, perform a second division with $b$ playing the role of $a$ and $%
r_{1} $ playing the role of $b$:
\[
b=q_{2}r_{1}+r_{2}, \quad 0 < r_2 < r_1,
\]
3) Next:
\[
r_{1}=q_{3}r_{2}+r_{3}, \quad 0 < r_3 < r_2.
\]
Continue in this way. When we finally obtain a remainder that
divides the
previous remainder, we are done: that final nonzero remainder is the \textit{%
g.c.d. }of $a$ and $b:$
\begin{eqnarray*}
r_{t} &=&q_{t+2}r_{t+1}+r_{t+2}, \\
r_{t+1} &=&q_{t+3}r_{t+2}.
\end{eqnarray*}
We obtain: $r_{t+2}=d=$\textit{g.c.d.}$(a,b\mathit{).}$

{\bf Example}. Find \textit{g.c.d}$.(128,24):$
\begin{eqnarray*}
128 &=&5\cdot 24+8, \\
24 &=&3\cdot 8
\end{eqnarray*}
We obtain that \textit{g.c.d}$.(128,24)=8.$

Let us prove that Euclid`s algorithm indeed gives the greatest
common divisor. Note first that $b>r_{1}>r_{2}>...$ is a sequence
of decreasing positive integers which can not be continued
indefinitely. Consequently Euclid`s algorithm must end.

Let us go up through out Euclid`s algorithm. $r_{t+2}=d$ divides $r_{t+1},$ $%
r_{t},...,r_{1},b,a.$ Thus $d$ is a common divisor of $a$ and $b.$

Now let $c$ be any common divisor of $a$ and $b.$ Go downward
through out Euclid`s algorithm. $c$ divides
$r_{1},r_{2},...,r_{t+2}=d.$ Thus $d,$ being a common divisor of
$a$ and $b,$ is divisible by any common divisor of these numbers.
Consequently $d$ is the greatest common divisor of $a$ and
$b.\Box$

Another (but similar) proof is based on the formula
\[
\mathit{g.c.d}.(qb+r,b)=\mathit{g.c.d}.(b,r).
\]
{\bf Corollary.} Note that from Euclid`s algorithm it follows
 (go up) that if $d=$\textit{%
g.c.d.}$(a,b\mathit{)}$ then there are integers $u$ and $v$ such
that
\begin{equation}
d=ua+vb.  \label{num2}
\end{equation}
In particular one has
\begin{equation}
ua\equiv d~(\text {mod}~ b)  \label{num2a}
\end{equation}

One can estimate the efficiency of Euclid`s algorithm. By
\textit{Lame`s theorem} the number of divisions required to find
the greatest common divisor of two integers is never greater that
five-times the number of digits in the smaller integer.

\textbf{Congruences.} An integer $a$ is \textit{congruent to }$b$ \textit{%
modulo }$m,$
\[
a\equiv b ~(\text{mod}~m)
\]
iff $m$ divides $(a-b).$ In this case $a=b+km$ where $k=0,\pm
1,\pm 2,...$.

{\bf Proposition}. Let us be given two integers $a$ and $m$. The
following are equivalent

(i) There exists $u$ such that $au\equiv 1~(\text{mod}~m).$

(ii) $\mathit{g.c.d}.(a,m)=1.$

\textbf{Proof.} From (i) it follows
\[
ab-mk=1.
\]
Therefore the $\mathit{g.c.d}.(a,m)=1,$ i.e. we get (ii).

Now if (ii) is valid then one has the relation (\ref{num2a}) for
$d=1,b=m$:
\[
au\equiv 1~(\text {mod}~m)
\]
which gives (i).$\Box$

Let us solve in integers the equation
\begin{equation}
ax\equiv c~(\text{mod}~ m)  \label{num3}
\end{equation}
We suppose that $\mathit{g.c.d}.(a,m)=1.$ Then by the previous
proposition there exists such $b$ that
\[
ab\equiv 1~(\text{mod}~m).
\]
Multiplying Eq (\ref{num3}) to $b$ we obtain the solution
\begin{equation}
x\equiv bc~(\text{mod}~m)  \label{num4}
\end{equation}
or more explicitly
\[
x=bc+km,\qquad k=0,\pm 1,\pm 2,...
\]
{\bf Exercise.} Find all of the solutions of the congruence
$$
3x\equiv 4~(\text{mod}~7).
$$

\textbf{Chinese Remainder Theorem.} Suppose there is a system of
congruences to different moduli:
\begin{eqnarray*}
x &\equiv &a_{1}~(\text{mod}~{m_1}), \\
x &\equiv &a_{2}~(\text{mod}~m_{2}), \\
&&... \\
x &\equiv &a_{t}~(\text{mod}~ m_{t})
\end{eqnarray*}
Suppose  $\mathit{g.c.d}.(m_{i},m_{j})=1$ for $i\neq j.$ Then
there exists a solution $x$ to all of the congruences, and any two
solutions are congruent to one another modulo
\[
M=m_{1}m_{2}...m_{t}.
\]
\textbf{Proof. }Let us denote $M_{i}=M/m_{i}.$ There exist $N_{i}$
such that
\[
M_{i}N_{i}\equiv 1~(\text{mod}~m_{i})
\]
Let us set
\[
x={\sum }_i a_{i}M_{i}N_{i}
\]
This is the solution. Indeed we have
\[
{\sum }_i a_{i}M_{i}N_{i}=a_{1}M_{1}M_{1}+...\equiv
a_{1}+a_{2}+...\equiv a_{1}~(\text{mod}~m_{1})
\]
and similarly for other congruences.$\Box$

We will need also

{\bf Fermat`s Little Theorem}. Let $p$ be a prime number. Any
integer $a$ satisfies
\[
a^{p}\equiv a~(\text{mod}~p)
\]
and any integer $a$ not divisible by $p$ satisfies
\[
a^{p-1}\equiv 1~(\text{mod}~p).
\]
\textbf{Proof.} Suppose $a$ is not divisible by $p$. Then
$\{0a,1a,2a,...,(p-1)a\}$ form a complete set of residues modulo
$p$, i.e. $\{a,2a,...,(p-1)a\} $ are a rearrangement of
$\{1,2,...,p-1\}$ when considered modulo $p$. Hence the product of
the numbers in the first sequence is congruent modulo $p$ to the
product of the members in the second sequence, i.e.
$$
a^{p-1}(p-1)\equiv (p-1)!~(\text{mod}~p)
$$
Thus $p$ divides $(p-1)(a^{p-1}-1)$. Since $(p-1)!$ is not
divisible by $p$, it should be that $p$ divides
$(a^{p-1}-1)$.$\Box$

{\bf The Euler function.}

The Euler function $\varphi(n)$ is the number of nonnegative
integers $a$ less then $n$ which are prime to $n$:
$$
\varphi(n)=\#\{0\leq a < n:g.c.d.(a,n)=1\}
$$
In particular $\varphi (1)=1,~\varphi (2)=1,...,\varphi
(6)=2,...$. One has $\varphi (p)=p-1$ for any prime $p$.

{\bf Exercise.} Prove: $\varphi (p^n)=p^n-p^{n-1}$ for any $n$ and
prime $p$.

The Euler function is multiplicative, meaning that
$$
\varphi (mn)=\varphi (m)\varphi (n)
$$
whenever $g.c.d.(m,n)=1$.

If
$$
n=p_1^{\alpha_1}p_2^{\alpha_2}...p_k^{\alpha_k}
$$
then
$$
\varphi (n)=n(1-\frac{1}{p_1})...(1-\frac{1}{p_k})
$$
In particular, if $n$ is the product of two primes, $n=pq$, then
$$
\varphi (n)=\varphi (p)\varphi (q)=(p-1)(q-1)
$$
There is the following generalization of Fermat`s Little Theorem.

{\bf Euler`s theorem.} If $g.c.d.(a,m)=1$ then
$$
a^{\varphi (m)}\equiv 1~(\text{mod}~m).
$$
{\bf Proof.} Let $r_1,r_2,...,r_{\varphi (m)}$ be classes of
integers relatively prime to $m$. Such a system is called a
reduced system of residues mod $m$. Then
$ar_1,ar_2,...,ar_{\varphi (m)}$ is another reduced system since
$g.c.d.(a,m)=1$. Therefore
$$ar_1\equiv r_{\pi (1)},~ar_2\equiv r_{\pi (2)},...,ar_{\varphi (m)}\equiv
r_{\pi (m)}~(\text{mod}~m)
$$
On multiplying these congruences, we get
$$
a^{\varphi (m)}r_1r_2...r_{\varphi (m)}\equiv r_1r_2...r_{\varphi
(m)}~(\text{mod}~m)
$$
Now since $r_1r_2...r_{\varphi (m)}$ is relatively prime to $m$
the theorem is proved.$\Box$

\section{Public Key Cryptography and RSA Cryptosystem}

First let us define some extra notions that we will use along with ones
defined in the previous sections.

{\bf Information channel} - a way to transmit information from one endpoint
to another.

{\bf Trusted channel} - an information channel where it is
believed that is impossible to eavesdrop the transmitted
information. For example military optical communication channels.

{\bf Public channel} - an information channel where the
transmitted information could be quite easily overheard. An
example is the Internet.

Let us introduce our main characters: Alice, Bob and Eve. Alice
wants to send ciphertext to Bob. Eve, the eavesdropper, wants to
catch the ciphertext and {\bf break} it, i.e. decipher without
knowing the deciphering key. In our scheme in order to produce a
ciphertext from the plaintext Alice has to have an {\it
enciphering key}. In turn, Bob to read (decipher) the Alice's
ciphertext needs a {\it deciphering key}. If Alice and Bob use a
{\it private key cryptosystem}, i.e. a cryptosystem where
enciphering and deciphering keys could be easily produced one from
another they come to the {\it key distribution problem}. Indeed
Alice and Bob should use a {\it trusted} channel to share the
keys.

From the first glance it seems to be impossible to get rid of the need of
the secret channel. However in 1976 W. Diffie and M. Hellman \cite{DH}
discovered a new
type of cryptosystem called {\it public key cryptosystem} where there is no
key distribution problem at all. A public key cryptosystem has the property
that having the enciphering key one cannot find the deciphering key without
a prohibitively lengthy computation. In other words the enciphering
function $f: \EuScript{P}\to\EuScript{C}$ is easy to compute if the
enciphering key $K_E$ is known, but it is very hard to compute the inverse
function $f^{-1}: \EuScript{C}\to\EuScript{P}$ without knowing the {\it
deciphering} key $K_D$ even having the {\it enciphering} key $K_E$.

One of the most widely used public key cryptosystem is RSA - a
cryptosystem named after the three inventors, Ron Rivest, Adi
Shamir, and Leonard Adleman \cite{RSA}. 
The RSA cryptosystem is based on the
fact that in order to factorise a big natural number with $N$
digits any classical computer needs at least a number of steps
that grows faster than any polynomial in $N$. Faithfully speaking
there is no rigorous proof of this fact but all known factoring
algorithms obey this fact.

Let us describe RSA cryptosystem in more detail. First we describe
the {\it protocol}, i.e. the steps our characters Alice and Bob
should perform in order to allow Alice send enciphered messages to
Bob. The mathematical basis of the RSA cryptosystem will be
described in the next section.

\subsection{The RSA Protocol}

The RSA protocol solves the following problem.
Bob wants to announce publicly a public key such that
Alice using this key will send to him an enciphering
message and nobody but Bob will be able to decipher it.
 

1. Bob generates public and private keys - each of them
is a pair of two natural numbers~- $(e,n)$ and $(d,n)$.
Here $K_e=(e,n)$ is the enciphering key (public) and $K_d=(d,n)$ is
the deciphering key (private).

In order to generate public and private keys Bob does the following:

\vspace{2mm}
~~~~~\parbox{.9\linewidth}{
a) Takes any two big prime numbers $p$ and $q$ and compute $n=pq$
and the value of the Euler function $\varphi (n)=(p-1)(q-1)$.
In modern cryptosystems one uses $\log p \approx \log q \approx 1000$.
}

\vspace{2mm}
~~~~~\parbox{.9\linewidth}{
b) Takes any $e < n$, such that $\gcd(e,\varphi(n))=1$.
}

\vspace{2mm}
~~~~~\parbox{.9\linewidth}{
c) Computes $d=e^{-1}\md{\varphi(n)}$, i.e. finds natural $d$ such that
\begin{equation}
\label{eq-xx}
ed\equiv 1\md{\varphi(n)},~1\leq d < \varphi(n)
\end{equation}
}
\vspace{2mm}

2. Bob sends a public key $(n,e)$ to Alice via a public channel.

3. Alice having Bob's public key $(n,e)$ and a plaintext $m$ (assume $m$ is
a natural number and $m<n$) that she wants to send to Bob computes
$$
c=m^e\md{n}
$$
and sends $c$ (ciphertext) to Bob.

4. When Bob receives $c$ from Alice he computes
$$
c^d\md{n}
$$
and gets the Alice's plaintext $m$, because $m=c^d\md{n}$

Nobody but Bob will be able to decipher Alice`s message.

\subsection{Mathematical Basis of the RSA Protocol}

In this section we will show why the RSA cryptosystem works. Then we will
discuss the {\it security} of the protocol, i.e. how hard for Eve, the
eavesdropper, to decipher the Alice's message without knowing the private
key.

If order to prove that RSA cryptosystem works we have to prove that the
computations that Bob does on the step d). of the protocol is inverse to the
computations that Alice does on the step c). That is
$$
c^d\equiv m\md{n}
$$
From (\ref{eq-xx}) we have
$$
ed=1+k\varphi(n),~k\in\mathbb{Z}
$$
We have
\begin{equation}
\label{eq-1x}
c^d=m^{ed}=m\cdot m^{k\varphi(n)}
\end{equation}
Finally using the Euler's theorem for the rhs of (\ref{eq-1x}) we obtain
$$
c^d\equiv m\md{n}. \Box
$$

~

Now let us investigate the security of the RSA cryptosystem. It seems to be
rather straightforward for Eve to obtain the Bob's {\it private} key having
his {\it public} key. The only thing she has to do is having $n$ and $e$
solve the congruence
$$
de\equiv 1\md{\varphi(n)},~1\leq d < \varphi(n)
$$
The problem that Eve would face here is to compute $\varphi(n).$
To this end she has to know $p$ and $q$, i.e. she has to solve
the factoring problem. The practical solution of this problem
is not possible with modern technology.
For a discussion of this problem see
for example \cite{IVol}.





\section{Shannon's Entropy and Mutual Information }

Here we  summarize some notions from information theory
\cite{OP,Ohy,May} used in quantum cryptography
 for the consideration
of security of quantum  cryptographic protocols.

Privacy is often expressed in terms of {\em Shannon's entropy}
or {\em mutual information}. Let $(\Omega,{\cal F},P)$
be a probability space and  
$X$, $Y$ and $Z$  three  random variables taking values
in a discrete set on the real line.  Let 
$p(x,y,z) = P(X = x \wedge Y = y \wedge Z
= z)$ is the joint distribution,
$p(x,y) = P(X =
x \wedge Y = y)$ is the marginal distribution,
$p(x|y) =
P(X = x\, | \, Y = y)$ is the conditional distribution,
and $p(x) = P(X = x)$,\ \ $p(y) = P(Y = y)$.

 The {\it Shannon entropy} of $X$ is given by
$$
H(X)  = - \sum_x p(x) \log p(x).
$$
The {\it mutual information} between $X$ and $Y$ is given by
$$
I(X\,;\,Y) = \sum_{x,y} p(x,y) \log\left(\frac{p(x,y)}{p(x)p(y)}\right).
$$
 {\it The conditional Shannon entropy of $X$ given $Y$ } is given by
$$
H(X\,|\,Y) = - \sum_{x,y} p(x,y) \log p(x|y).
$$
One has
$$I(X\,;\,Y) = H(X) - H(X\,|\,Y) = H(Y) - H(Y\,|\,X).
$$ 
The {\it conditional mutual information} between $X$ and $Y$ given
$Z$ is
\begin{eqnarray*}
I(X\,;\,Y\,|\,Z) & = & \sum_{x,y,z} p(x,y,z) \log
\left(\frac{p(x,y|z)} {p(x|z) p(y|z)} \right) 
\end{eqnarray*}

Quantum entropy of an observable $A$ in the state $\rho$ is defined by
\begin{equation}
\label{et}
H(A,\rho)=-\sum_i p(i,\rho) \log p(i,\rho)
\end{equation}
where $p( \cdot ,\rho)$ is the probability 
distribution of an observable $A$
in the state $\rho$. If the state $\rho$ is pure, i.e.
$\rho=|\varphi\>\<\varphi|$, where $\varphi$
is a unit vector in a Hilbert space, one can rewrite (\ref{et})
as
\begin{equation}
\label{epure}
H(A,\varphi)=-\sum_i |\<\xi_i|\varphi\>|^2 \log |\<\xi_i|\varphi\>|^2
\end{equation}
where $\{|\xi_i\>\}$ is an orthonormal
basis consisting from  eigenvectors of the observable $A$.

\section{Entropic Uncertainty Relations}
\label{ur}

The fundamental Heisenberg uncertainty relation
is a particular case of the Robertson inequality
$$
\Delta(A,\psi)\Delta(B,\psi)\geq \frac{1}{2}
|\<\psi|[A,B]\psi\>|
$$
where $A$ and $B$ are two observables
and 
$$
\Delta(A,\psi)=\sqrt{\<\psi|(A-\<\psi|A\psi\>)^2\psi\>}
$$
Here we discuss a generalization of the uncertainty
relation which uses the notions of entropy
and mutual information.

{\bf Theorem 1.}
For any nondegenerate observables $A$ and $B$ in the finite dimensional
Hilbert space the {\it entropic uncertainty relation} holds \cite{MU,OP}
\begin{equation}
\label{etrne}
H(A,\rho)+H(B,\rho) \geq -2\log c
\end{equation}
where $c$ is defined as the maximum possible
overlap of the eigenstates of $A$ and $B$
\begin{equation}
\label{cd}
c\equiv\max\limits_{a,b} |\<a|b\>|
\end{equation}
Here $\{|a\>\}$ and $\{|b\>\}$ are orthonormal bases
consisting from eigenvectors of $A$ and $B$ respectively.

One can check that for any nondegenerate observable $A$ in $N$-dimensional
Hilbert space there exists an upper bound on the entropy
\begin{equation}
\label{etneq}
H(A,\rho) \leq \log N
\end{equation}

Let us illustrate the entropic uncertainty relation on a simple
spin-$\frac{1}{2}$ particle. Taking Pauli matrices
\begin{equation}
\label{pauli}
\sigma_x=\left(
    \begin{array}{cc}0 & 1\\ 1 & 0
    \end{array}
\right),~~
\sigma_z=\left(
    \begin{array}{cc}1 & 0\\ 0 & -1
    \end{array}
\right)
\end{equation}
as an observables with eigenstates
\begin{equation}
\label{basis}
h_1=\frac{1}{\sqrt 2}
\left(
    \begin{array}{c}1\\1
    \end{array}
    \right),~~
h_2=\frac{1}{\sqrt 2}
\left(
    \begin{array}{c}1\\-1
    \end{array}
    \right),~~
e_1=\left(
    \begin{array}{c}0\\1
    \end{array}
    \right),~~
e_2=\left(
    \begin{array}{c}1\\0
    \end{array}
    \right)
\end{equation}
we compute $c=1/\sqrt{2}$. Now taking $2$ as a base of the logarithm, the
relation (\ref{etrne}) states that for any unit vector
$\varphi\in\mathbb{C}^2$
it holds
\begin{equation}
\sum_{i=1,2} ( |\<e_i|\varphi\>|^2 \log |\<e_i|\varphi\>|^2 +
             |\<h_i|\varphi\>|^2 \log |\<h_i|\varphi\>|^2) \leq -1
\end{equation}

Now we will formulate the uncertainty relation
using the mutual information.
Consider a quantum system which is described by density  
operator $\rho_i$
with probability $p_i$. Then the density operator of the whole ensemble
${\cal E}=\{\rho_i\}$ of all possible states of the system is given by
$$
\rho=\sum_i p_i \rho_i
$$
The mutual information corresponding to a measurement of an
observable $A$ is given by
$$
I(A,{\cal E})=H(A,\rho)-\sum_i p_i H(A,\rho_i)
$$

From (\ref{etrne}) using (\ref{etneq}) one can obtain 
the following theorem
(information exclusion relation \cite{Hal})

{\bf Theorem 2.} Let $A$ and $B$ be arbitrary 
observables in $N$-dimensional
Hilbert space, then
$$
I(A,{\cal E})+I(B,{\cal E}) \leq 2\log Nc
$$
where $c$ is defined by (\ref{cd}).

\section{The No Cloning Theorem}\label{nc}

The eavesdropper, Eve, wants to have a perfect copy
of Alice`s message. However 
 Wootters and Zurek \cite{WZ} proved
  that perfect
copying is impossible in the quantum world. 

It is instructive to start with the following

{\bf Proposition.} If ${\cal H}$
is a Hilbert space and $\phi_0$ is a vector from
${\cal H}$ then there is no a linear map 
$M : {\cal H}\otimes {\cal H}\to
{\cal H}\otimes {\cal H}$ with the property
$M(\psi\otimes \phi_0)=
 \psi\otimes \psi$ for any  $\psi$.

{\bf Proof.} Indeed we would have 
$$M(2\psi\otimes \phi_0)=
2\psi\otimes 2\psi=
 4\psi\otimes \psi$$
But because of linearity
 we should have 
$$M(2\psi\otimes \phi_0)=
 2M(\psi\otimes \phi_0)=2\psi\otimes \psi$$
This contradiction proves the claim. Now let us prove the
no cloning theorem.
 
{\bf Theorem.} Let ${\cal H}$ and ${\cal K}$ be two
Hilbert spaces, $\dim {\cal H} \geq 2.$ Let $M$ be a 
a linear map (copy machine) 
$$
M : {\cal H}\otimes {\cal H}\otimes {\cal K}\to
{\cal H}\otimes {\cal H}\otimes {\cal K}
$$
with the property
$$
M(\psi\otimes \phi_0\otimes\xi_0)=
 \psi\otimes \psi\otimes\eta_{\psi}
$$
 for any $\psi\in {\cal H}$ and some nonzero
vectors $\psi_0\in {\cal H}$ and $\xi_0\in {\cal K}$ 
where $\eta_{\psi}\in {\cal K}$ can depend
on $\psi$. Then $M$ is a trivial map, $M=0$ (i.e. $\eta_{\psi}=0$ for any
$\psi$).

{\bf Proof.} Let $\{e_i\}$ be an orthonormal basis in ${\cal H}$.
We have 
$$
M(e_i\otimes \phi_0\otimes\xi_0)=
 e_i\otimes e_i\otimes\eta_i
$$
where $\eta_i$ are some vectors in ${\cal K}$.
To prove the theorem we prove
that $\eta_i=0.$
If $i\neq j$ then $(e_i +e_j)/\sqrt 2$ is 
a unit vector (here we use that $\dim {\cal H} \geq 2$). 
We have the equality
$$
\frac{1}{\sqrt 2}(e_i +e_j)\otimes\phi_0\otimes\xi_0
=\frac{1}{\sqrt 2}e_i\otimes\phi_0\otimes\xi_0
+\frac{1}{\sqrt 2} e_j\otimes\phi_0\otimes\xi_0
$$
Let us apply the map $M$ to both sides of this equality.
Then we get
\begin{equation}
\label{ct}
\frac{1}{\sqrt 2}(e_i+e_j)\otimes\frac{1}{\sqrt 2}(e_i +e_j)
\otimes\eta_{ij}=
\frac{1}{\sqrt 2}e_i\otimes\frac{1}{\sqrt 2}e_i\otimes\eta_i
+\frac{1}{\sqrt 2}e_j\otimes\frac{1}{\sqrt 2}e_j\otimes\eta_j
\end{equation}
where $\eta_{ij}$ is a vector
in ${\cal K}$.
We can rewrite (\ref{ct}) as
$$
e_i\otimes e_i\otimes (\eta_{ij}-\eta_i)
+e_i\otimes e_j\otimes \eta_{ij}
+e_j\otimes e_i\otimes \eta_{ij}
+e_j\otimes e_j\otimes (\eta_{ij}-\eta_j)=0
$$
Now  taking into account
that $e_i$ and $e_j$ belong to a basis in ${\cal H}$
we get 
$$
\eta_{ij}-\eta_i=0,~~\eta_{ij}=0,~~\eta_{ij}-\eta_j=0
$$
Hence $\eta_i=0$ for any $i$  and Theorem is proved.

{\bf Remark.} If $\dim {\cal H} = 1,$ i.e. ${\cal H}=\mathbb{C}$,
then Theorem is not valid. For $\phi_0 =1$ and $\psi\in\mathbb{C}$
one can set  $M(\psi\xi_0)=\psi\xi_0=\psi^2\eta_{\psi}$
where $\eta_{\psi}=\xi_0/\psi$ for $\psi\neq 0.$
 
We proved that Eve can not get a perfect 
quantum copy because perfect quantum
copy machines can not exist. The possibility to copy classical information
is one of the most crucial features of information needed for
eavesdropping. The quantum no cloning theorem prevents Eve from perfect
eavesdropping, and hence makes quantum cryptography potentially secure.

Note however that though there is no a perfect quantum cloning machine
but there are cloning machines that achieve the optimal approximate cloning
transformation compatible with the no cloning theorem, see \cite{GM,CIV}.
 
\section{The BB84 Quantum Cryptographic Protocol}

Quantum cryptographic protocols differ from the classical ones in that
their security is based on the laws of quantum mechanics, rather than the
conjectured computational difficulty of certain functions. In this section
we will describe the Bennett and Brassard (BB84)  quantum cryptographic
protocol \cite{BB}.

\subsection{The BB84 Protocol}

First let us describe the physical devices used by Alice and Bob.

Alice has a {\bf photon emitter} - a device which is capable to
emit single photons that are linear polarized in one of four
directions.
The polarizations are described by the four unit vectors
in $\mathbb{C}^2$ here they are $e_1, e_2, h_1, h_2$ given in (\ref{basis}).
We will call the polarizations
vertical, horizontal, diagonal, anti-diagonal ones and denote them
respectively (~$\mid$~,~---~,~$\diagdown$~,~$\diagup$~).  
We have two bases in $\mathbb{C}^2$. One basis, $G_z=\{e_1,e_2\}$,
describes the vertical and horizontal polarizations.
Another basis, $G_x=\{h_1,h_2\}$, describes the diagonal and anti-diagonal
polarizations.
Note that one has
\begin{equation}
\label{basort}
|(e_i,h_j)|=1/\sqrt 2, ~~ i,j=1,2
\end{equation} 
Bases with such a property are called conjugate. Note also that the vectors
$e_1,e_2$ from the basis $G_z$ and $h_1,h_2$ from the basis $G_x$ are the
eigenvectors of the Pauli matrices $\sigma_z$ and $\sigma_x$ respectively,
see (\ref{pauli}). 

Bob has a {\bf photon detector} - a device that detects
single photons in one of the two bases.

Alice can send photons emitted by the photon emitter to Bob and
Bob detects the photons with the photon detector. 
~

{\bf The Protocol.}

1. Alice chooses a random polarization basis and prepares photons with a
random polarization that belongs to the chosen basis. She sends the photons
to Bob.

2. For each photon Bob chooses at random which polarization basis he will
use, and measures the polarization of the photon. (If Bob chooses the same
basis as Alice he can for sure identify the polarization of the photon).

3. Alice and Bob use the public channel to compare the polarization bases
they used. They keep only the polarization data for which the polarization
bases are the same. In the absence of errors and eavesdropping these data
should be the same on both sides, it is called a {\it raw key}.

4. At the last step Alice and Bob use methods of classical information
theory to check whether their raw keys are the same.  For example, they
choose a random subset of the raw key and compare it using the public
channel. They compute {\it the error rate} 
(that is, the fraction of data for
which their values disagree). If the error rate is unreasonably high -
above, say, $10\%$ - they 
abort the protocol and may be try again later. If the
error rate is not that high they could use error correction codes.

As a result of the protocol Alice and Bob share the same random data. This
data could now be used as a private key in the symmetric cryptosystems.

Instead of polarized photons one can use any two level quantum system.
One can consider also a generalized quantum key distribution protocol
using a $d$-dimensional Hilbert space 
with $k$  bases, each basis has $d$  states, \cite{Bru,BKBGC,CBKG,BP}.

\subsection{BB84 Security}
\label{bb84s}

In transmitting information, 
there are always some errors and Alice and Bob must apply
some classical information processing protocols
 to improve their data . They can use
{\it error correction} to obtain identical keys and
{\it privacy amplification}  to obtain a
secret key. To solve the problem of eavesdropping 
one has to find a protocol which, assuming
that Alice and Bob can only measure the error rate of the received data,
either provides Alice and Bob with a  secure key, or aborts the
protocol and tells the parties that the key distribution has failed. 
There are various eavesdropping problems, depending  in particular
on the technological power which Eve could have   and
on the assumed fidelity of Alice and Bob's devices, 
\cite{GRTZ,CIV,FGG}. 

There is a simple eavesdropping strategy, called intercept-resend.
Eve measures each qubit in one of the two basis and resends to Bob
a qubit in the state corresponding to the result of her measurement.
This attack belongs to the class of the so called individual
attacks.
In this way Eve will get $50\%$ information. However Alice and Bob
can detect the actions of Eve because they will have $25\%$ of errors
in their sifted key. But it would be not so easy
to detect eavesdropping  if Eve applies
the intercept-resend
strategy to only a fraction of the Alice`s sending.

In this case one can use methods
of classical cryptography. We suppose that 
once Alice, Bob and Eve have made
their measurements, they will get classical random variables
$\alpha,\beta$ and $\epsilon$ respectively, with a joint probability
distribution $p(x,y,z)$. Let $I(\alpha ,\beta)$ be the mutual
information of Alice and Bob and $I(\alpha ,\epsilon)$
and $I(\beta ,\epsilon)$ the mutual information
of Alice and Eve and Bob and Eve respectively.
Intuitively, it is clear 
that only if Bob has more information on Alice`s bits
then Eve then it could  be possible to establish
a secret key between Alice and Bob. In fact one can prove
(see \cite{CK,GRTZ}) the following 

{\bf Theorem 1.}  Alice and Bob can
establish a secret key (using error correction and privacy amplification)
if, and only if 
$$
I(\alpha ,\beta)\geq I(\alpha, \epsilon)~~\text{or}~~
I(\alpha ,\beta)\geq I(\beta, \epsilon).
$$

Let $D$ be the error rate. Then
one can prove that the BB84 protocol is secure
against individual attacks if one has the following  bound
$$
 D<D_0\equiv\frac{1-1/
\sqrt{2}}{2}\approx 15\%
$$
There have been discussed also more general coherent
or joint attacks when Eve measures several qubits simultaneously. 
An important problem of the eavesdropping analysis 
is to find quantum cryptosystems
for which one can prove its {\it ultimate security}. Ultimate
security means that the security is guaranteed against 
the whole  class of
eavesdropping attacks, even if Eve uses 
any conceivable technology of future.

We  assume that Eve has perfect technology which is only limited by the
laws of quantum mechanics.  This means she can use any unitary
transformation between any number of qubits and  an arbitrary auxiliary
system. But Eve is not allowed to come to Alice's  lab and read all her
data.

\subsection{Ultimate Security Proofs}

Main ideas on how to prove security of BB84 protocol were presented by D.
Mayers \cite{May} in 1996. The security issues are considered in recent
papers \cite{May,LC,MY,BBB,SP}, \cite{Bru}-\cite{ZLG}. We describe here a
simple and general  method proposed in  \cite{GRTZ,BKBGC,CBKG}. The method
is based on  Theorem 1 from Sect. \ref{bb84s} on classical cryptography
and on Theorem 2 from Sect. \ref{ur} on information uncertainty relations.

The argument  runs as follows. Suppose Alice sends out a large number of
qubits and   Bob receives $n$ of them  in the correct basis. The relevant
Hilbert space dimension is then $2^n$. Let us re-label the bases used for
each of the $n$ qubits in such a way that Alice used $n$ times the $x$-basis. 
Hence, Bob's observable is the $n$-time tensor product
$\sigma_x\otimes\ldots\otimes\sigma_x$. Since Eve had no way to know the
correct bases, her optimal information on the correct ones is precisely the
same as her optimal information on the incorrect ones. Hence one can bound
her information assuming she measures $\sigma_z\otimes...\otimes\sigma_z$.
Therefore $c=2^{-n/2}$ and  Theorem 2 from Sect. \ref{ur} implies:
\begin{equation}
\label{ii}
I(\alpha,\epsilon)+I(\alpha,\beta)\le 2\log_2(2^n2^{-n/2})=n
\end{equation}
 Next, combining the bound
(\ref{ii}) with Theorem 1 from Sect. \ref{bb84s}, one deduces that a secret key is achievable
if $I(\alpha,\beta)\ge n/2$. Using
$$I(\alpha,\beta)=n\left(1-D\log_2 D-(1-D)\log_2(1-D)\right)
$$ one
obtains the sufficient condition on the error rate $D$:
$$
-D\log_2 D-(1-D)\log_2(1-D) \le \frac{1}{2}
$$
i.e. $D\le 11\%$. This bound was obtained in Mayers proof
(after improvement by P. Shor and J. Preskill\cite{SP}). It is 
compatible with the $15\%$ bound found for individual attacks.

One can argue, however, that previous arguments lead
in fact to another
result: $c=2^{-n/4}$.
Indeed,
Bob's observable is the $n$-time tensor product
$
\sigma_x \otimes ..... \otimes \sigma_x.
$
Now, since Eve had no way to know the correct
basis it was assumed that she measures
$
\sigma_z \otimes ..... \otimes \sigma_z.
$
However it seems  if Eve does not know the correct
basis then her observables $\sigma_i$ will be complementary observables
to $\sigma_x$ only in the half of cases. In the other half of
cases her observables $\sigma_i$ will be the same as Bob's, 
i.e. $\sigma_x$.
Therefore one gets:
$
c=(1/\sqrt 2)^{n/2}=2^{-n/4}.
$
This  leads to a lower error rate, instead of $11\%$ one gets $4\%$.

\section{The EPRBE Quantum Cryptographic Protocol}

\subsection{Quantum Nonlocality and Cryptography}

Bell's theorem~\cite{Bel} states that there are quantum correlation
functions that can not be represented as classical correlation functions of
separated  random variables. It has been interpreted as incompatibility of
the requirement of locality with the statistical predictions of quantum
mechanics~\cite{Bel}. For a recent discussion of Bell's theorem see, for
example ~\cite{AfrSel} - ~\cite{Vol3} and references 
therein. It is now widely
accepted, as a result of Bell's theorem and related experiments, that
"local realism" must be rejected.

Bell's theorem constitutes an important part in quantum
cryptography~\cite{Eke}. It is now generally accepted that techniques of
quantum cryptography can allow secure communications between distant
parties  . The promise of some secure cryptographic
quantum key distribution schemes is based on the use of  quantum
entanglement in the spin space and  on quantum no-cloning theorem. An
important contribution of quantum cryptography is a mechanism for detecting
eavesdropping.

Let us stress that  the very formulation of the 
problem of locality in quantum
mechanics is based on ascribing a special role to the position in ordinary
three-dimensional space. However the space
dependence of the wave function is neglected in   many
discussions of the
problem of locality in relation to Bell's inequalities. Actually it is the
space part of the wave function which is relevant to the consideration of
the problem of locality.

It was  pointed out in \cite{Vol1} 
that the space part of the wave function
leads to an extra factor in quantum correlation which 
changes the Bell equation. It was
suggested
a criterion of locality (or nonlocality) of quantum theory
in a realist model of hidden variables. In particular predictions
of quantum mechanics can be consistent with Bell's inequalities
for some Gaussian wave functions.

If one neglects the space part of
the wave function in a cryptographic scheme
then such a scheme  could  be insecure in
the real three-dimensional space.

We will discuss how one can
try to improve the security of quantum 
cryptography schemes in space  by
using  a special preparation of 
the space part of the wave function,
see \cite{Vol2}.

\subsection{Bell's Inequalities}

In the  presentation of Bell's theorem we will 
follow ~\cite{Vol1} where one
can find also more references, see \cite{Vol3}
for more details. The mathematical formulation of Bell's
theorem reads:
\begin{equation}
\cos(\alpha - \beta)\neq E\xi_{\alpha}\eta_{\beta}
\label{eq:cos}
\end{equation}
where $\xi_{\alpha}$ and $\eta_{\beta}$ are two random
processes such that $|\xi_{\alpha}|\leq 1$,~
$|\eta_{\beta}|\leq 1$ and $E$ is the expectation. Let us
discuss in more details the physical interpretation of this result.

Consider a pair of spin one-half particles formed in the singlet spin state
and moving freely towards two detectors (Alice and Bob).  If one neglects
the space part of the wave function  then the quantum mechanical
correlation of two spins in the singlet state $\psi_{spin}$ is
\begin{equation}
 D_{spin}(a,b)=\left<\psi_{spin}|\sigma\cdot a \otimes\sigma\cdot
b|\psi_{spin}\right>=-a\cdot b
\label{eq:eqn1}
\end{equation}
Here $a$ and $b$ are two unit vectors in three-dimensional space,
$\sigma=(\sigma_1,\sigma_2,\sigma_3)$ are the Pauli matrices and
$$
\psi_{spin}=\frac{1}{\sqrt 2}
\left(\left(
    \begin{array}{c}0\\1
    \end{array}
    \right)
\otimes \left(
    \begin{array}{c}1\\
    0\end{array}
    \right)
-\left(
    \begin{array}{c}1\\
    0\end{array}
    \right)
\otimes
\left(
    \begin{array}{c}0\\
    1\end{array}
    \right)
\right)
$$

Bell's theorem states that the function $ D_{spin}(a,b)$
Eq.~(\ref{eq:eqn1}) can not be represented in the form
\begin{equation}
\label{eq:eqn2} P(a,b)=\int \xi (a,\lambda) \eta (b,\lambda)
d\rho(\lambda)
\end{equation}
i.e.
\begin{equation}
\label{eq:Ab}
D_{spin}(a,b)\neq P(a,b)
\end{equation}
Here $ \xi (a,\lambda)$ and $  \eta(b,\lambda)$ are random  fields on the
sphere, $|\xi (a,\lambda)|\leq 1$,~  $ | \eta (b,\lambda)|\leq 1$ and
$d\rho(\lambda)$ is a positive probability measure,  $ \int
d\rho(\lambda)=1$. The parameters $\lambda$ are interpreted as hidden
variables in a realist theory. It is clear that Eq.~(\ref{eq:Ab}) can be
reduced to Eq.~(\ref{eq:cos}).

One has the following  Bell-Clauser-Horn-Shimony-Holt (CHSH) inequality
\begin{equation}
\label{eq:eqn3}
 |P(a,b)-P(a,b')+P(a',b)+P(a',b')|\leq 2
\end{equation}
From the other hand there are such vectors $(ab=a'b=a'b'=-ab'=\sqrt{2}/2)$
for which one has
\begin{equation}
\label{eq:eqn4}
 | D_{spin}(a,b)- D_{spin}(a,b')+
 D_{spin}(a',b)+ D_{spin}(a',b')|=2\sqrt{2}
\end{equation}
Therefore if one supposes that $ D_{spin}(a,b)=P(a,b)$ then one gets the
contradiction.

It will be shown below that if one takes into account the space part of the
wave function then  the quantum correlation in the simplest case will take
the form $g \cos (\alpha - \beta)$ instead of just  $\cos
(\alpha - \beta)$ where the parameter $g$ describes the
location of the system in space and time. In this case one can get the
representation \cite{Vol1}
\begin{equation}
g\cos(\alpha - \beta)= E\xi_{\alpha}\eta_{\beta}
\label{eq:gcos}
\end{equation}
if $g$ is small enough (see below). The factor $g$ gives a contribution to
visibility or efficiency of detectors that are used in the phenomenological
description of detectors.

\subsection {Localized Detectors}

In the previous section the space part of the wave function of the
particles was neglected. However exactly the space part is relevant to the
discussion of locality. The complete wave function is $\psi
=(\psi_{\alpha\beta}({\bf r}_1,{\bf r}_2))$ where $\alpha$ and
$\beta $ are spinor indices and ${\bf r}_1$ and ${\bf r}_2$
are vectors in three-dimensional space.

We suppose that Alice and Bob have detectors which are located within the
two localized regions ${\cal O}_A$ and ${\cal O}_B$ respectively, well
separated from one another.

Quantum correlation describing the measurements of spins by Alice and Bob
at their localized  detectors is

\begin{equation}
\label{eq:eqn6}
 G(a,{\cal O}_A,b,{\cal O}_B)=\left<\psi| \sigma\cdot a   P_{{\cal O}_A}
 \otimes  \sigma\cdot b  P_{{\cal O}_B} |\psi\right>
\end{equation}
Here $P_{{\cal O}}$ is the projection operator onto the region ${\cal O}$.

Let us consider the case when the wave function has the form of the product
of the spin function and the space function $\psi=\psi_{spin}\phi({\bf
r}_1,{\bf r}_2)$. Then one has
\begin{equation}
\label{eq:eqn7}
 G(a,{\cal O}_A,b,{\cal O}_B)=g ({\cal O}_A,{\cal O}_B)
  D_{spin}(a,b)
\end{equation}
where the function
\begin{equation}
\label{eq:eqn8}
 g ({\cal O}_A,{\cal O}_B)=\int_{{\cal O}_A \times {\cal O}_B}|\phi({\bf
 r}_1,{\bf
 r}_2)|^2 d{\bf r}_1d{\bf r}_2
\end{equation}
describes correlation of particles in space. It is the probability to find
one particle in the region ${\cal O}_A$ and another particle in the region
${\cal O}_B$.

One has
\begin{equation}
\label{eq:eqn8g} 0\leq g ({\cal O}_A,{\cal O}_B)\leq 1
\end{equation}

{\bf Remark.} In relativistic quantum field theory there is no nonzero
strictly localized projection operator that annihilates the vacuum. It is
a consequence of the Reeh-Schlieder theorem.  Therefore, apparently, the
function $ g ({\cal O}_A,{\cal O}_B)$ should be always strictly smaller
than 1. 

Now one inquires whether one can write the representation
\begin{equation}
\label{eq:eqn9}
 g({\cal O}_A,{\cal O}_B)D_{spin}(a,b)=\int \xi (a,{\cal O}_A,\lambda)
 \eta (b,{\cal O}_B,\lambda) d\rho(\lambda)
\end{equation}

Note that if we are interested in the conditional probability of
finding the projection of spin along  vector $a$ for the particle
1  in the region ${\cal O}_A$ and the projection of spin along the
vector $b$ for the particle 2 in the region ${\cal O}_B$   then we
have to divide both sides of Eq.~(\ref{eq:eqn9}) to $g({\cal
O}_A,{\cal O}_B)$.

The factor $g$ is important. In particular one can write the following
representation \cite{VV} for $0\leq g\leq 1/2$:
\begin{equation}
\label{eq:gek}
g\cos(\alpha-\beta)=
\int_0^{2\pi}\sqrt {2g}\cos(\alpha-\lambda) \sqrt {2g}\cos(\beta-\lambda)
 \frac{d\lambda}{2\pi}
\end{equation}

Let us now apply these considerations to quantum cryptography.

\subsection {The EPRBE Quantum Key Distribution}

Ekert \cite{Eke} showed that one can use the Einstein-Podolsky-Rosen 
correlations to establish
a secret random key between two parties ("Alice" and "Bob"). Bell's
inequalities are used to check the presence of an intermediate eavesdropper
("Eve"). We will call this method
the Einstein-Podolsky-Rosen-Bell-Ekert
(EPRBE) quantum cryptographic protocol.
 There are two stages to the EPRBE protocol, the first stage over a
quantum channel, the second over a public channel.

The quantum channel consists of a source that emits pairs of spin one-half
particles, in a singlet state. The particles fly apart towards Alice and
Bob, who, after the particles have separated, perform measurements on spin
components along one of three directions, given by unit vectors $a$ and
$b$. In the second stage Alice and Bob communicate over a public
channel.They announce in public the orientation of the detectors they have
chosen for particular measurements. Then they divide the measurement
results into two separate groups: a first group for which they used
different orientation of  the detectors, and a second group for which they
used the same orientation of the detectors. Now Alice and Bob can reveal
publicly the results they obtained but within the first group of
measurements only. This allows them, by using Bell's inequality, to
establish the presence of an eavesdropper (Eve). The results of the second
group of measurements can be converted into a secret key. One supposes that
Eve  has a detector which is located within the region ${\cal O}_E$ and she
is described by hidden variables $\lambda$.

We will interpret Eve as  a  hidden variable in a realist theory and  will
study whether the quantum correlation Eq.~(\ref{eq:eqn7}) can be
represented in the form Eq.~(\ref{eq:eqn2}).  From (\ref{eq:eqn3}),
(\ref{eq:eqn4}) and (\ref{eq:eqn9}) one can see that if the following
inequality
\begin{equation}
\label{eq:eqn10}
 g ({\cal O}_A,{\cal O}_B)\leq 1/\sqrt 2
\end{equation}
is valid for  regions  ${\cal O}_A$ and ${\cal O}_B$ which are well
separated from one another then there is no violation of the CHSH
inequalities (\ref{eq:eqn3}) and therefore Alice and Bob can not detect the
presence of an eavesdropper. On the other side, if for a pair of well
separated regions ${\cal O}_A$ and ${\cal O}_B$ one has
\begin{equation}
\label{eq:eqn11}
 g ({\cal O}_A,{\cal O}_B) > 1/\sqrt 2
\end{equation}
then it could be a  violation of the realist locality in these regions for
a given state. Then, in principle, one can hope to detect an eavesdropper
in these circumstances.

Note that if we set $ g({\cal O}_A,{\cal O}_B)=1$ in (\ref{eq:eqn9}) as it
was done in the original proof of Bell's theorem, then it means we did a
special preparation of the states of particles to be completely localized
inside of detectors. There exist such well localized states (see however
the previous Remark) but there exist also another states, with the wave
functions which are not very well localized inside the detectors, and still
particles in such states are also observed in detectors. The fact that a
particle is observed inside the detector does not mean, of course, that its
wave function is strictly localized inside the detector before the
measurement.  Actually one has  to perform a thorough investigation of the
preparation and the  evolution of our entangled states in space and time if
one needs to estimate the function $ g({\cal O}_A,{\cal O}_B)$.

\subsection {Gaussian Wave Functions}

Now let us consider the criterion of locality for Gaussian wave
functions. We will show that with a reasonable accuracy there is
no violation of locality in this case. Let us take the wave
function $\phi$ of the form $\phi=\psi_{1}({\bf r}_1)\psi_{2}({\bf
r}_2)$  where the individual wave functions have the moduli
\begin{equation}
\label{eq:eqn12}
|\psi_{1}({\bf
 r})|^2 =(\frac{m^2}{2\pi})^{3/2}e^{-m^2{\bf r}^2/2},~~
|\psi_{2}({\bf
 r})|^2 =(\frac{m^2}{2\pi})^{3/2}e^{-m^2 ({\bf r}- {\bf l})^2/2}
\end{equation}
We suppose that  the length of the vector ${\bf l}$ is much larger
than $1/m$. We can make measurements of  $P_{{\cal O}_A}$ and
$P_{{\cal O}_B}$ for any  well separated regions  ${\cal O}_A$ and
${\cal O}_B$. Let us suppose a rather nonfavorite case for the
criterion of locality when the wave  functions of the particles
are almost localized inside the regions ${\cal O}_A$ and ${\cal
O}_B$ respectively. In such  a case the function $g ({\cal
O}_A,{\cal O}_B)$ can take values near its maximum. We suppose
that the region ${\cal O}_A$ is given by $|r_i|<1/m, {\bf
r}=(r_1,r_2,r_3) $ and the region ${\cal O}_B$ is obtained from
${\cal O}_A$ by translation on ${\bf l}$. Hence $\psi_{1}({\bf
r}_1)$ is a Gaussian function with modules appreciably different
from zero only in ${\cal O}_A$ and similarly $\psi_{2}({\bf r}_2)$
is localized in the region ${\cal O}_B$. Then we have
\begin{equation}
\label{eq:eqn13}
 g ({\cal O}_A,{\cal O}_B)=\left(\frac{1}{\sqrt {2\pi}}\int_{-1}^1
e^{-x^2/2}dx\right)^6
\end{equation}
One can estimate (\ref{eq:eqn13}) as
\begin{equation}
\label{eq:eqn14}
 g ({\cal O}_A,{\cal O}_B)< \left(\frac{2}{\pi}\right)^3
\end{equation}
which is smaller than $1/2$. Therefore the locality criterion
(\ref{eq:eqn10}) is satisfied in this case.

Let us remind that there is a well known effect of expansion of
wave packets due to the free time evolution. If $\epsilon$ is the
characteristic length of the Gaussian wave packet describing a
particle of mass $M$ at time $t=0$  then at time $t$ the
characteristic length  $\epsilon_t$ will be
\begin{equation}
\label{eq:epsil}
\epsilon_t=\epsilon\sqrt{1+\frac{\hbar^2t^2}{M^2\epsilon^4}}.
\end{equation}

It tends to $(\hbar/M\epsilon)t$ as $t\to\infty$. Therefore the
locality criterion is always satisfied for nonrelativistic
particles if regions ${\cal O}_A$ and ${\cal O}_B$ are far enough
from each other. 

\section{Conclusions}

In quantum cryptography there are many interesting open problems which
require  further investigations. In quantum cryptographic protocols with
two entangled photons (such as the EPRBE protocol) to detect the
eavesdropper's presence by using Bell's inequality we have to estimate the
function  $ g({\cal O}_A,{\cal O}_B)$. In order to increase the
detectability of the eavesdropper one has to do a thorough investigation of
the process of preparation  of the entangled state and then its  evolution
in space and time towards Alice and Bob. One has to develop a proof of the
security of such a protocol.

In the previous section Eve was interpreted as an abstract hidden variable.
However one can assume that more information about Eve is available. In
particular one can assume that she is located somewhere in space in a
region ${\cal O}_E.$ It seems that one has to study a generalization of the
function  $ g({\cal O}_A,{\cal O}_B)$, which depends not only on the Alice
and Bob locations ${\cal O}_A$ and $ {\cal O}_B$ but also on Eve's location
${\cal O}_E$. Then one can try to find a strategy which leads to an optimal
value of this function.

In quantum cryptographic protocols with single photons
(such as the BB84 protocol) further investigation of 
the security  under various types of attacks, including
attacks from real space, would be desirable.


\section*{Acknowledgments}

This work was supported in part by RFFI 99-0100866
 and by INTAS 99-00545 grants.


\end{document}